\documentclass[aps, prd, preprint, amsfonts,
  amssymb, amsmath, showpacs, a4paper,nofootinbib,superscriptaddress]{revtex4-1}

\usepackage{slashed}
\usepackage{braket, bm}
\usepackage{setspace}

\begin{document}

\raggedbottom

\count\footins=1000

\title{Discrete spacetime symmetries, second quantization, and inner products in a non-Hermitian Dirac fermionic field theory}

\author{Jean Alexandre}
\email{jean.alexandre@kcl.ac.uk}
\affiliation{Department of Physics, King's College London,\\ 
London WC2R 2LS, United Kingdom}

\author{John Ellis}
\email{john.ellis@cern.ch}
\affiliation{Department of Physics, King's College London,\\ 
London WC2R 2LS, United Kingdom}
\affiliation{National Institute of Chemical Physics \& Biophysics, R\"avala 10, 10143 Tallinn, Estonia}
\affiliation{Theoretical Physics Department, CERN, CH-1211 Geneva 23, Switzerland}

\author{Peter Millington}
\email{peter.millington@manchester.ac.uk}
\affiliation{Department of Physics and Astronomy, University of Manchester,\\ Manchester M13 9PL, United Kingdom}
\affiliation{School of Physics and Astronomy, University of Nottingham,\\ Nottingham NG7 2RD, United Kingdom\vspace{1cm}}

\begin{abstract}
{\small
We extend to a non-Hermitian fermionic quantum field theory with $\mathcal{PT}$ symmetry our previous discussion of second quantization, discrete symmetry transformations, and inner products in a scalar field theory [\href{https://doi.org/10.1103/PhysRevD.102.125030}{Phys.\ Rev.\ D {\bf 102} (2020) 125030}]. For illustration, we consider a prototype model containing a single Dirac fermion with a parity-odd, anti-Hermitian mass term.  In the phase of unbroken $\mathcal{PT}$ symmetry, this Dirac fermion model is equivalent to a Hermitian theory under a similarity transformation, with the non-Hermitian nature of the 
model residing only in the spinor structure, whereas the algebra
of the creation and annihilation operators is just that of a
Hermitian theory.
~~\\
~~\\
KCL-PH-TH/2021-79, CERN-TH-2021-172}
\\
\footnotesize{This is an author-prepared post-print of \href{https://doi.org/10.1103/PhysRevD.106.065003}{Phys.\ Rev.\ D {\bf 106} (2022) 065003}, published by the American Physical Society under the terms of the \href{https://creativecommons.org/licenses/by/4.0/}{CC BY 4.0} license (funded by SCOAP\textsuperscript{3}).}
\end{abstract}

\maketitle


\section{Introduction}

Quantum systems with non-Hermitian Hamiltonians that possess $\mathcal{PT}$ symmetry, where $\mathcal{P}$ and $\mathcal{T}$ denote parity and time-reversal, have real energy levels and exhibit unitary time evolution~\cite{Bender:1998ke}. These~\cite{Bender:2005tb} and the wider class of pseudo-Hermitian quantum theories~\cite{Mostafazadeh:2001jk, Mostafazadeh:2001nr, Mostafazadeh:2002id} have attracted growing interest in recent years, driven also by the applications of such theories in many different fields, including photonics~\cite{Longhi, Longhi2, El-Ganainy} and phase transitions~\cite{Ashida, Matsumoto:2019are}. In view of this increasing importance, it is desirable to formulate more carefully $\mathcal{PT}$-symmetric quantum field theories, verifying the arguments for their consistency, and analyzing the structures of their Fock spaces, their discrete symmetries and inner products.~\footnote{See, e.g., Ref.~\cite{Mannheim:2017apd} for a discussion of the inner products in $\mathcal{PT}$-symmetric quantum mechanics.} We recently analyzed these issues in a prototype minimal field theory containing a pair of scalar bosons that are free apart from non-Hermitian mixing~\cite{AEMB}. 

In this paper, we extend our analysis of this bosonic field theory to a
minimal non-Hermitian quantum field theory with a single Dirac spin-1/2 fermion
that possesses $\mathcal{PT}$ symmetry at the classical and quantum levels. We 
formulate the discrete symmetries of this Dirac model and discuss candidate inner products in
Fock space.

A prototype non-Hermitian theory with a single fermion flavour $\psi$ was originally studied in Ref.~\cite{BJR}, where
the anti-Hermitian but $\mathcal{PT}$-symmetric fermion mass term $\mu\bar\psi\gamma^5\psi$ was considered. The tachyonic ($\mathcal{PT}$-broken) regimes of this model were studied in Refs.~\cite{Jentschura:2012rd, Jentschura:2012vp, Jentschura:2012aw, Jentschura:2013nsa}. It was shown in Ref.~\cite{AB} that this model corresponds to chiralities with different current densities, and similar behaviour has been reproduced in a $1+1$-dimensional lattice model~\cite{Chernodub:2017lmx}.
The corresponding gauged fermion model was studied in Ref.~\cite{ABM}, wherein it was shown how the prototype model can be obtained from a non-Hermitian Higgs-Yukawa theory (see also Ref.~\cite{Alexandre:2017fpq}), discussed further in the context of the type-I seesaw in Ref.~\cite{Mishra:2018aej}. An extension to include four-fermion interactions was studied in Ref.~\cite{Beygi:2019qab}. The supersymmetric version of this model was analyzed in Ref.~\cite{AEM4}. Reference~\cite{Ohlsson:2015xsa} studied a fermionic model with a single wavefunction
and oscillations between different energy states, which is possible with a non-Hermitian Hamiltonian 
only if the latter has complex coefficients, and its application to neutrino mixing and oscillations (first suggested in Ref.~\cite{Jones-Smith:2009qeu}) was considered. 
It was found in Ref.~\cite{AEMB} that
the apparent puzzles concerning the positivity of transition probabilities and unitarity found in Refs.~\cite{Ohlsson:2019noy, Ohlsson:2020gxx} (cf.~Ref.~\cite{Ohlsson:2020idi} for the phase of broken $\mathcal{PT}$ symmetry)
are resolved in a bosonic model
by recalling that these requirements apply only to asymptotic states, which are not problematic in these respects.

As we discussed in Ref.~\cite{AEMB}, whereas
the inner product in the quantum Fock space is unique 
in a bosonic theory described by a Hermitian Hamiltonian, 
theories with non-Hermitian Hamiltonians can be formulated
using different definitions of the inner product. However,
energy eigenstates are not orthogonal with respect to the
usual Dirac inner product, and the norm with respect to the $\mathcal{PT}$ inner product is not positive-definite.
This issue in the formulation of the Fock space in a non-Hermitian model
with $\mathcal{PT}$ symmetry may be resolved by
introducing a discrete symmetry $\mathcal{C'}$,\footnote{This was introduced in the quantum mechanics case in Ref.~\cite{Bender:2002vv} as the $\mathcal{C}$ symmetry. We refer to it as the $\mathcal{C'}$ symmetry to distinguish it from charge conjugation.} defined previously in quantum-mechanical systems and in a bosonic $\mathcal{PT}$-symmetric
model~\cite{AEMB}, and using the $\mathcal{C'PT}$ 
inner product that yields a positive-definite norm.~\footnote{An alternative approach has been proposed in Ref.~\cite{Mannheim:2017apd}.}

For definiteness, we frame the discussions that follow in the context of the prototype non-Hermitian but $\mathcal{PT}$-symmetric 
non-interacting Dirac fermion field theory of Ref.~\cite{BJR}, formulated in $3+1$ dimensional Minkowski spacetime. This model contains a single Dirac fermion  and is described by a
non-Hermitian Lagrangian with real parameters.  It comprises four degrees of freedom, the minimal number needed to realise a non-Hermitian, 
$\mathcal{PT}$-symmetric field theory with real Lagrangian parameters. We show that the non-Hermiticity of this model resides only in the spinor structure, such that the Fock space remains that of a Hermitian theory. In this way, and while the $\mathcal{C}'$ transformation can still be constructed, we argue that there is no subtlety to defining the inner product between single-particle states, so long as one works with the correct canonical conjugate spinor field operators.

The layout of our paper is as follows.  First, in Section~\ref{sec:Dirac},
we introduce the Dirac fermion field theory that we study, then reviewing in Section~\ref{sec:discretesymms}
its discrete $\mathcal{P}$ and $\mathcal{T}$ symmetries at the classical level. We introduce the
two-component Weyl spinor formulation of the theory in Section~\ref{sec:twocomponent},
and present a discussion of quantization in four-component notation in the Dirac basis in Section~\ref{sec:quant}. Next, we discuss a useful matrix theory in Section~\ref{sec:matrix}
and display a similarity transformation to a Hermitian theory in Section~\ref{sec:similarity}.
We revisit in Section~\ref{sec:Fock} the discrete symmetries, including the
$\mathcal{C}'$ symmetry~\cite{Bender:2002vv}, and discuss inner products in Section~\ref{sec:inner}.
Finally, we summarize our conclusions in Section~\ref{sec:conx}.


\section{Dirac model}
\label{sec:Dirac}

The fermionic model that we study is composed of a single Dirac fermion with both a Hermitian and an anti-Hermitian mass term. The $c$-number Lagrangian is~\cite{BJR}
\begin{equation}
\label{eq:DiracLag}
\mathcal{L}
=\bar{\psi}i\slashed{\partial}\psi-m\bar{\psi}\psi-\mu\bar{\psi}\gamma^5\psi~,
\end{equation}
where $m$ and $\mu$ are both real mass parameters, and $\bar{\psi}\equiv\psi^{\dag}\gamma^0$ is the usual Dirac-conjugate spinor. The squared eigenmasses are
\begin{equation}
M^2=m^2-\mu^2~,
\end{equation}
which are real in the $\mathcal{PT}$-symmetric regime $\mu^2< m^2$. The fifth gamma matrix, which appears explicitly in Eq.~\eqref{eq:DiracLag}, is given in the Dirac basis by
\begin{equation}
    \gamma^5=\left(\gamma^{5}\right)^{\dag}=i\gamma^0\gamma^1\gamma^2\gamma^3=\begin{pmatrix} 0_{2} & \mathbb{I}_{2} \\ \mathbb{I}_{2} & 0_{2}\end{pmatrix}~,
\end{equation}
where $0_{2}$ and $\mathbb{I}_2$ are the $2\times 2$ zero and unit matrices, respectively. At the so-called exceptional points (unique to non-Hermitian theories), when $\mu=\pm m$, the mass matrix becomes defective, and we lose a degree of freedom:~the theory reduces to one of either a massless left- or a right-chiral Weyl fermion (see, e.g., Refs.~\cite{ABM, Alexandre:2017fpq}). In fact, this model is most easily studied in two-component notation in the Weyl basis, as we describe in Section~\ref{sec:twocomponent}. We note that the conserved current corresponding to the global $U(1)$ symmetry of the Lagrangian~\eqref{eq:DiracLag}, which was originally derived in Ref.~\cite{AB}, has non-trivial properties under improper Lorentz transformations~\cite{AM}.

The non-Hermiticity of the Lagrangian in Eq.~\eqref{eq:DiracLag} means that the variations of the corresponding action with respect to $\psi$ and $\bar{\psi}$ do not yield identical equations of motion, since they differ by $\mu\to -\mu$. However, as explained in Ref.~\cite{AMS}, we are free to choose either of these equations of motion, since physical observables depend only on $\mu^2$ and are therefore independent of this sign. Choosing the equations of motion obtained by varying with respect to $\bar{\psi}$, we obtain the following non-Hermitian Dirac equation
\begin{equation}
    \label{eq:Diraceq1}
    i\slashed{\partial}\psi-m\psi-\mu\gamma^5\psi=0~.
\end{equation}
However, as we will see in the next Section, and as was first established for scalar quantum field theories in Ref.~\cite{AEMB}, there exists an alternative Lagrangian for this model that yields the same equations of motion without the need for the above prescription for obtaining the dynamics by varying with respect to $\psi$ or $\bar \psi$, but not both, as we now describe.


\section{Discrete symmetries}
\label{sec:discretesymms}

We recall that the classical Lagrangian \eqref{eq:DiracLag} is $\mathcal{PT}$ symmetric under the following naive transformations of the $c$-number fields~\cite{AMS}:
\begin{subequations}
\begin{align}
\mathcal{P}:\qquad &\psi(t,\mathbf{x})\to \psi'(t,-\mathbf{x})=P\psi(t,\mathbf{x})~,\nonumber\\
&\bar{\psi}(t,\mathbf{x})\to \bar{\psi}'(t,-\mathbf{x})=\bar{\psi}(t,\mathbf{x})P~,\\
\mathcal{T}:\qquad &\psi(t,\mathbf{x})\to \psi'(-t,\mathbf{x})=T\psi^*(t,\mathbf{x})~,\nonumber\\
&\bar{\psi}(t,\mathbf{x})\to \bar{\psi}'(-t,\mathbf{x})=\bar{\psi}^*(t,\mathbf{x})T~,
\end{align}
\end{subequations}
where $P=\gamma^0$ and $T=i\gamma^1\gamma^3$. However, the Lagrangian in Eq.~\eqref{eq:DiracLag} is not invariant under parity, since the action of parity changes the sign of the anti-Hermitian term. Parity therefore maps between the two possible choices of equation of motion obtainable from Eq.~\eqref{eq:DiracLag}.

Taking this into account and following Ref.~\cite{AEMB}, it is more convenient to work with the Lagrangian~\footnote{Equivalently, we could work with the Lagrangian 
    $\tilde{\tilde{\mathcal{L}}}=\psi^{\dag}\gamma^0i\slashed{\partial}\tilde{\psi}-m\psi^{\dag}\gamma^0\tilde{\psi}-\mu\psi^{\dag}\gamma^0\gamma^5\tilde{\psi}$~,
which gives equations of motion consistent with fixing the dynamics by varying Eq.~\eqref{eq:DiracLag} with respect to $\psi$, i.e., the alternative choice to that made in the main text.}
\begin{equation}
    \label{eq:DiracLtilde}
    \tilde{\mathcal{L}}
    =\tilde{\psi}^{\dag}\gamma^0i\slashed{\partial}\psi-m\tilde{\psi}^{\dag}\gamma^0\psi-\mu\tilde{\psi}^{\dag}\gamma^0\gamma^5\psi
\end{equation}
and its Hermitian conjugate, where the equation of motion of the tilded field $\tilde{\psi}$ is the parity conjugate of that of the untilded field $\psi$, i.e., if $\psi$ satisfies Eq.~\eqref{eq:Diraceq1} then
\begin{equation}
    \label{eq:Diraceq2}
    i\slashed{\partial}\tilde{\psi}-m\tilde{\psi}+\mu\gamma^5\tilde{\psi}=0~.
\end{equation}
The equations of motion obtained from varying the corresponding action with respect to $\psi$ and $\tilde{\psi}^{\dag}$ are now mutually consistent.

The consistent discrete spacetime transformations of the $c$-number spinors are then
\begin{subequations}
\label{eq:tildePTtransformations}
\begin{align}
\mathcal{P}:\qquad &\psi(t,\mathbf{x})\to \psi'(t,-\mathbf{x})=P\tilde{\psi}(t,\mathbf{x})~,\nonumber\\
&\psi^{\dag}(t,\mathbf{x})\to \psi^{\dag\prime}(t,-\mathbf{x})=\tilde{\psi}^{\dag}(t,\mathbf{x})P~,\\
\mathcal{T}:\qquad &\psi(t,\mathbf{x})\to \psi'(-t,\mathbf{x})=T\psi^*(t,\mathbf{x})~,\nonumber\\
&\psi^{\dag}(t,\mathbf{x})\to \psi^{\dag\prime}(-t,\mathbf{x})=\psi^{\mathsf{T}}(t,\mathbf{x})T~,
\end{align}
\end{subequations}
and we see that the classical Lagrangian~\eqref{eq:DiracLtilde} is $\mathcal{PT}$ symmetric.


\section{Two-component Weyl basis}
\label{sec:twocomponent}

It will prove convenient to consider the decomposition of the Dirac four-spinor in terms of two two-component Weyl spinors $\chi_1$ and $\chi_2$ in the chiral (Weyl) basis. In this basis, the four-component Dirac spinor can be written as
\begin{equation}
\label{chichitildec}
    \psi=\begin{pmatrix}
    \chi_{2,\alpha}\\ \tilde{\chi}_1^{\dag\dot{\alpha}}
    \end{pmatrix}~,
\end{equation}
where $\alpha$ and $\dot{\alpha}$ are spinor indices. We note that the four-spinor necessarily involves the untilded Weyl spinor $\chi_2$ and the Hermitian conjugate of the tilded Weyl spinor $\tilde{\chi}_1$, due to the requirement that the four-spinor $\psi$ evolves with respect to the Hamiltonian $H$. This implies that both Weyl components must evolve with the \emph{same} Hamiltonian, whereas the Weyl spinor and its Hermitian conjugate evolve with $H$ and $H^{\dag}\neq H$, respectively.

In terms of the two-component $c$-number Weyl spinors, the Lagrangian takes the form
\begin{equation}
    \label{eq:WeylLagcnumber}
    \tilde{\mathcal{L}}=i\tilde{\chi}^{\dag}_{i,\dot{\alpha}}\bar{\sigma}^{\nu\dot{\alpha}\beta}\partial_{\nu}\chi_{i,\beta}-(m-\mu)\chi_1^{\alpha}\chi_{2,\alpha}-(m+\mu)\tilde{\chi}_{2,\dot{\alpha}}^{\dag}\tilde{\chi}_1^{\dag\dot{\alpha}}~,
\end{equation}
where $i\in\{1,2\}$, $\bar{\sigma}^{\mu}\equiv(\sigma^0,-\bm{\sigma})$ and $\sigma^{\mu}\equiv(\sigma^0,\bm{\sigma})$, 
with the Pauli matrices $\bm{\sigma}\equiv(\sigma^1,\sigma^2,\sigma^3)$.
The resulting equations of motion are
\begin{subequations}
\begin{align}
    i\bar{\sigma}^{\nu\dot{\alpha}\beta}\partial_{\nu}\chi_{i,\beta}-(m+\mu)\tilde{\chi}_{\slashed{i}}^{\dag\dot{\alpha}}=0~,\\
    i\bar{\sigma}^{\nu\dot{\alpha}\beta}\partial_{\nu}\tilde{\chi}_{i,\beta}-(m-\mu)\chi_{\slashed{i}}^{\dag\dot{\alpha}}=0~,
\end{align}
\end{subequations}
along with their Hermitian conjugates. Herein, $\slashed{i}=2$ if $i=1$, and $\slashed{i}=1$ if $i=2$. Notice that tilde conjugation takes $\mu\leftrightarrow-\mu$ but not $\bar{\sigma}\leftrightarrow \sigma$.

The corresponding momentum-space Dirac equations for the two independent $c$-number two-spinors $x$ and $y$ are
\begin{subequations}
\begin{align}
    (\bar{\sigma}\cdot p)^{\dot{\alpha}\beta} \, x_{\beta}(\mathbf{p},s)-(m+\mu) \, \tilde{y}^{\dag\dot{\alpha}}(\mathbf{p},s)=0~,\\
    (\bar{\sigma}\cdot p)^{\dot{\alpha}\beta} \, \tilde{x}_{\beta}(\mathbf{p},s)-(m-\mu) \, y^{\dag\dot{\alpha}}(\mathbf{p},s)=0~,\\
    (\bar{\sigma}\cdot p)^{\dot{\alpha}\beta} \, y_{\beta}(\mathbf{p},s)+(m+\mu) \, \tilde{x}^{\dag\dot{\alpha}}(\mathbf{p},s)=0~,\\
    (\bar{\sigma}\cdot p)^{\dot{\alpha}\beta} \, \tilde{y}_{\beta}(\mathbf{p},s)+(m-\mu) \, x^{\dag\dot{\alpha}}(\mathbf{p},s)=0~,
\end{align}
\end{subequations}
along with their Hermitian conjugates, wherein we have followed the notation of Ref.~\cite{Dreiner:2008tw}, with the exception that we denote the helicity index by $s=\pm$. The explicit expressions for the two-component Weyl spinors are
\begin{subequations}
\label{eq:twospinors}
\begin{align}
x_{\alpha}(\mathbf{p},s)&=\sqrt[4]{\frac{m+\mu}{m-\mu}}(\sqrt{\sigma\cdot p})_{\alpha}^{\phantom{\alpha}\beta} \, u_{s,\beta}(\mathbf{s})~,\\
\tilde{x}_{\alpha}(\mathbf{p},s)&=\sqrt[4]{\frac{m-\mu}{m+\mu}}(\sqrt{\sigma\cdot p})_{\alpha}^{\phantom{\alpha}\beta} \, u_{s,\beta}(\mathbf{s})~,\\
y^{\dag\dot{\alpha}}(\mathbf{p},s)&=\sqrt[4]{\frac{m+\mu}{m-\mu}}(\sqrt{\bar{\sigma}\cdot p})^{\dot{\alpha}}_{\phantom{\dot{\alpha}}\dot{\beta}} \, u_{s}^{\dot{\beta}}(\mathbf{s})~,\\
\tilde{y}^{\dag\dot{\alpha}}(\mathbf{p},s)&=\sqrt[4]{\frac{m-\mu}{m+\mu}}(\sqrt{\bar{\sigma}\cdot p})^{\dot{\alpha}}_{\phantom{\dot{\alpha}}\dot{\beta}} \, u_{s}^{\dot{\beta}}(\mathbf{s})~,
\end{align}
\end{subequations}
which reduce to the standard expressions in the Hermitian limit $\mu\to 0$ (see Ref.~\cite{Dreiner:2008tw}). Herein,
\begin{equation}
\mathbf{s} \equiv \mathbf{p}/|\mathbf{p}|\equiv (\sin \theta\cos \phi,\sin\theta\sin\phi,\cos\theta)    
\end{equation}
in spherical polar coordinates,  and the $u_s(\mathbf{s})$ are the two-spinors\footnote{Herein, we employ conventions similar to those used in Appendix B of Ref.~\cite{BhupalDev:2014pfm}.}
\begin{equation}
    u_s(\mathbf{s})=\begin{cases} \begin{pmatrix} \cos \frac{\theta}{2}e^{-i\frac{\phi}{2}}\\ \sin\frac{\theta}{2}e^{i\frac{\phi}{2}}\end{pmatrix}~,&\qquad s=+\\ i\begin{pmatrix} -\sin \frac{\theta}{2}e^{-i\frac{\phi}{2}}\\ \cos\frac{\theta}{2}e^{i\frac{\phi}{2}}\end{pmatrix}~,&\qquad s=-\end{cases}~,
\end{equation}
satisfying
\begin{equation}
    \label{eq:us_identity}
    \frac{\bm{\sigma}\cdot\mathbf{p}}{|\mathbf{p}|} \, u_s(\pm\mathbf{s})=\pm \, s \, u_s(\pm \mathbf{s})~.
\end{equation}
We note also that
\begin{equation}
x_{\alpha}(\mathbf{p},s)=\sqrt{\frac{m+\mu}{m-\mu}} \, \tilde{x}_{\alpha}(\mathbf{p},s)~,\qquad 
y^{\dag\dot{\alpha}}(\mathbf{p},s)=\sqrt{\frac{m+\mu}{m-\mu}} \, \tilde{y}^{\dag\dot{\alpha}}(\mathbf{p},s)~,
\end{equation}
and that the usual relationship between the $x$ and $y$ spinors persists:
\begin{equation}
    x(\mathbf{p},s)=-sy(\mathbf{p},-s)~,\qquad x^{\dag}(\mathbf{p},s)=-sy^{\dag}(\mathbf{p},-s)~.
\end{equation}

It can easily be checked that the spin sums for conjugate pairs of two-spinors have the usual forms~\cite{Dreiner:2008tw}:
\begin{subequations}
\label{eq:twocomponent_spin_sums}
\begin{align}
\sum_s x_{\alpha}(\mathbf{p},s)\tilde{x}_{\dot{\beta}}^{\dag}(\mathbf{p},s)&=(\sigma\cdot p)_{\alpha\dot{\beta}}~,\qquad &\sum_s y_{\alpha}(\mathbf{p},s)\tilde{y}^{\dag}_{\dot{\beta}}(\mathbf{p},s)&=(\sigma\cdot p)_{\alpha\dot{\beta}}~,\\
\sum_s y^{\dag\dot{\alpha}}(\mathbf{p},s)\tilde{y}^{\beta}(\mathbf{p},s)&=(\bar{\sigma}\cdot p)^{\dot{\alpha}\beta}~,\qquad &\sum_s x^{\dag\dot{\alpha}}(\mathbf{p},s)\tilde{x}^{\beta}(\mathbf{p},s)&=(\bar{\sigma}\cdot p)^{\dot{\alpha}\beta}~,\\
\sum_s x_{\alpha}(\mathbf{p},s)\tilde{y}^{\beta}(\mathbf{p},s)&=M\delta_{\alpha}^{\phantom{\alpha}\beta}~,\qquad &\sum_s y_{\alpha}(\mathbf{p},s)\tilde{x}^{\beta}(\mathbf{p},s)&=-M\delta_{\alpha}^{\phantom{\alpha}\beta}~,\\
\sum_s y^{\dag\dot{\alpha}}(\mathbf{p},s)\tilde{x}^{\dag}_{\dot{\beta}}(\mathbf{p},s)&=M\delta^{\dot{\alpha}}_{\phantom{\dot{\alpha}}\dot{\beta}}~,\qquad
&\sum_s x^{\dag\dot{\alpha}}(\mathbf{p},s)\tilde{y}^{\dag}_{\dot{\beta}}(\mathbf{p},s)&=-M\delta^{\dot{\alpha}}_{\phantom{\dot{\alpha}}\dot{\beta}}~,
\end{align}
\end{subequations}
along with the corresponding tilde-conjugated expressions, where $M=\sqrt{m^2-\mu^2}$ and we have used
\begin{equation}
    \sum_s u_s(\mathbf{s})u_s^{\dag}(\mathbf{s})= \sigma^0~.
\end{equation}

Given the expression for the four-spinor [Eq.~\eqref{chichitildec}] and its transformation properties [Eq.~\eqref{eq:tildePTtransformations}], we see that the parity and time-reversal transformations for the $c$-number two-spinors are as follows:
\begin{subequations}
\begin{align}
\mathcal{P}:\qquad &\chi_i(t,\mathbf{x})\to \chi_i'(t,-\mathbf{x})=\tilde{\chi}_{\slashed{i}}^{\dag}(t,\mathbf{x})~,\nonumber\\
&\tilde{\chi}_i(t,\mathbf{x})\to \tilde{\chi}_i'(t,-\mathbf{x})=\chi_{\slashed{i}}^{\dag}(t,\mathbf{x})~,\\
\mathcal{T}:\qquad &\chi_i(t,\mathbf{x})\to \chi_i'(-t,\mathbf{x})=T\chi_i^*(t,\mathbf{x})~,\nonumber\\
&\tilde{\chi}_i(t,\mathbf{x})\to \tilde{\chi}_i'(-t,\mathbf{x})=T\tilde{\chi}_i^*(t,\mathbf{x})~,
\end{align}
\end{subequations}
where $T=i\sigma^1\bar{\sigma}^3$ in the two-component basis. We can confirm that the Lagrangian is indeed $\mathcal{PT}$ symmetric.

Turning our attention to second quantization, we can decompose the two-component quantum fields as follows:
\begin{subequations}
\label{eq:DiracWeylops}
\begin{align}
\hat{\chi}_{1,\alpha}(x)&=\sum_{s=\pm}\int_{\mathbf{p}}\frac{1}{\sqrt{2E_{\mathbf{p}}}}\left[\hat{d}_{\mathbf{p},s}(0)x_{\alpha}(\mathbf{p},s)e^{-ip\cdot x}+\hat{b}^{\dag}_{\mathbf{p},s}(0)y_{\alpha}(\mathbf{p},s)e^{ip\cdot x}\right]~,\\
\check{\chi}_{1,\alpha}(x)&=\sum_{s=\pm}\int_{\mathbf{p}}\frac{1}{\sqrt{2E_{\mathbf{p}}}}\left[\hat{d}_{\mathbf{p},s}(0)\tilde{x}_{\alpha}(\mathbf{p},s)e^{-ip\cdot x}+\hat{b}^{\dag}_{\mathbf{p},s}(0)\tilde{y}_{\alpha}(\mathbf{p},s)e^{ip\cdot x}\right]~,\\
\hat{\chi}_{2,\alpha}(x)&=\sum_{s=\pm}\int_{\mathbf{p}}\frac{1}{\sqrt{2E_{\mathbf{p}}}}\left[\hat{b}_{\mathbf{p},s}(0)x_{\alpha}(\mathbf{p},s)e^{-ip\cdot x}+\hat{d}^{\dag}_{\mathbf{p},s}(0)y_{\alpha}(\mathbf{p},s)e^{ip\cdot x}\right]~,\\
\check{\chi}_{2,\alpha}(x)&=\sum_{s=\pm}\int_{\mathbf{p}}\frac{1}{\sqrt{2E_{\mathbf{p}}}}\left[\hat{b}_{\mathbf{p},s}(0)\tilde{x}_{\alpha}(\mathbf{p},s)e^{-ip\cdot x}+\hat{d}^{\dag}_{\mathbf{p},s}(0)\tilde{y}_{\alpha}(\mathbf{p},s)e^{ip\cdot x}\right]~,
\end{align}
\end{subequations}
and similarly for their Hermitian conjugates, where $p^0=E_{\mathbf{p}}=\sqrt{\mathbf{p}^2+M^2}$. The hatted ($\hat{\ }$) field operators correspond to the untilded $c$-number fields, and the checked ($\check{\ }$) operators correspond to the tilded $c$-number fields. The need to introduce these two distinct sets of field operators follows from the requirement that canonical-conjugate variables must both evolve with the \emph{same} Hamiltonian $\hat{H}$ (see Ref.~\cite{AEMB} and the earlier discussion in this Section). Whereas $\hat{\chi}_{i}$ and $\hat{\chi}^{\dag}_i$ evolve respectively with $\hat{H}$ and $\hat{H}^{\dag}\neq \hat{H}$, and therefore cannot be canonical-conjugate variables, both $\hat{\chi}_{i}$ and $\check{\chi}^{\dag}_i$ evolve with $\hat{H}$.

In terms of the spinors in Eq.~\eqref{eq:DiracWeylops}, the second-quantized version of the classical Lagrangian in Eq.~\eqref{eq:WeylLagcnumber} is
\begin{equation}
    \label{eq:DiracLagWeyl}
    \hat{\mathcal{L}}=i\check{\chi}^{\dag}_{i,\dot{\alpha}}\bar{\sigma}^{\nu\dot{\alpha}\beta}\partial_{\nu}\hat{\chi}_{i,\beta}-(m-\mu)\hat{\chi}_1^{\alpha}\hat{\chi}_{2,\alpha}-(m+\mu)\check{\chi}_{2,\dot{\alpha}}^{\dag}\check{\chi}_1^{\dag\dot{\alpha}}~.
\end{equation}
Varying with respect to either variable, i.e., $\hat{\chi}_i$ or $\check{\chi}^{\dag}_i$, we obtain mutually consistent equations of motion.\footnote{We emphasize that the definition of hatted versus checked operators is a choice:~interchanging their definitions would lead to an equivalent description.}

The creation and annihilation operators satisfy the equal-time anticommutation relations
\begin{align}
    \left\{\hat{b}_{\mathbf{p},s}(t),\hat{b}^{\dag}_{\mathbf{p}',s'}(t)\right\}=\left\{\hat{d}_{\mathbf{p},s}(t),\hat{d}^{\dag}_{\mathbf{p}',s'}(t)\right\}=(2\pi)^3\delta_{ss'}\delta^{3}(\mathbf{p}-\mathbf{p}')~.
\end{align}
This algebra is consistent with the canonical commutation relations of the field operators
\begin{equation}
    \left\{\hat{\chi}_i(t,\mathbf{x}),\check{\chi}^{\dag}_j(t,\mathbf{y})\right\}=\left\{\check{\chi}_i(t,\mathbf{x}),\hat{\chi}^{\dag}_j(t,\mathbf{y})\right\}=\delta_{ij}\delta^3(\mathbf{x}-\mathbf{y})~.
\end{equation}
Notice that, unlike the scalar example in Ref.~\cite{AEMB}, there is no need to introduce checked creation and annihilation operators $\check{b}_{\mathbf{p},s}(t)$ and $\check{d}_{\mathbf{p},s}(t)$, since there is only a single squared energy eigenvalue $E^2_{\mathbf{p}}=\mathbf{p}^2+M^2$.  Instead, for this non-Hermitian Dirac fermion model, the non-Hermiticity resides only in the spinor structure, which we take into account by introducing the tilded and untilded spinors, defined in this and the next Section.  With regard the algebra of the creation and annihilation operators, this proceeds as per the Hermitian case, and there is no subtlety to defining the inner product of Fock states (see Section~\ref{sec:inner}), which is just the usual Dirac inner product with respect to Hermitian conjugation.


\section{Four-component Dirac basis}
\label{sec:quant}

Before discussing the non-Hermitian structure of this model in detail, we include for completeness a discussion of quantization in four-component notation in the Dirac basis.

Given the equation of motion \eqref{eq:Diraceq1}, the Dirac equations for the momentum-space four-spinors $u(\mathbf{p},s)$ and $v(\mathbf{p},s)$ are
\begin{subequations}
\begin{align}
    \left(\slashed{p}-m-\mu\gamma^5\right) u(\mathbf{p},s)&=0~,\\
    \left(\slashed{p}+m+\mu\gamma^5\right)v(\mathbf{p},s)&=0~,
    \end{align} 
\end{subequations}
where $s=\pm$ again indicates the helicity. The solutions in the Dirac basis are
\begin{subequations}
\label{eq:fourspinors}
\begin{align}
u(\mathbf{p},s)&=\begin{pmatrix}\left[ \xi_+(E+M)^{1/2}-s\xi_-(E-M)^{1/2}\right]u_s(\mathbf{s})\\\left[ -\xi_-(E+M)^{1/2}+s\xi_+(E-M)^{1/2}\right]u_s(\mathbf{s})\end{pmatrix}~,\\
v(\mathbf{p},s)&=
\begin{pmatrix}s\left[ \xi_-(E+M)^{1/2}+s\xi_+(E-M)^{1/2}\right]u_{-s}(\mathbf{s})\\-s\left[ \xi_+(E+M)^{1/2}+s\xi_-(E-M)^{1/2}\right]u_{-s}(\mathbf{s})\end{pmatrix}~,
\end{align}
\end{subequations}
where $M=\sqrt{m^2-\mu^2}$, and we have defined
\begin{equation}
    \label{eq:xipm_def}
    \xi_{\pm}\equiv \frac{1}{2}\left[\sqrt[4]{\frac{m+\mu}{m-\mu}}\pm\sqrt[4]{\frac{m-\mu}{m+\mu}}\right]
\end{equation}
and set overall complex phases to zero for brevity. It is easy to check that
the four-spinors in Eq.~\eqref{eq:fourspinors} reduce to the more familiar Hermitian expressions in the limit $\mu\to 0$ [cf.~Eq.~\eqref{eq:Hermitian_fourspinors}]. We also introduce the tilded spinors
\begin{subequations}
\label{eq:fourspinors_tilded}
\begin{align}
\tilde{u}(\mathbf{p},s)&=\begin{pmatrix}\left[ \xi_+(E+M)^{1/2}+s\xi_-(E-M)^{1/2}\right]u_s(\mathbf{s})\\\left[ \xi_-(E+M)^{1/2}+s\xi_+(E-M)^{1/2}\right]u_s(\mathbf{s})\end{pmatrix}~,\\
\tilde{v}(\mathbf{p},s)&=
\begin{pmatrix}s\left[ -\xi_-(E+M)^{1/2}+s\xi_+(E-M)^{1/2}\right]u_{-s}(\mathbf{s})\\-s\left[ \xi_+(E+M)^{1/2}-s\xi_-(E-M)^{1/2}\right]u_{-s}(\mathbf{s})\end{pmatrix}~,
\end{align}
\end{subequations}
differing by $\mu\to-\mu$ (i.e., $\xi_-\to-\xi_-$) and satisfying
\begin{subequations}
\begin{align}
    \left(\slashed{p}-m+\mu\gamma^5\right)\tilde{u}(\mathbf{p},s)&=0~,\\
    \left(\slashed{p}+m-\mu\gamma^5\right)\tilde{v}(\mathbf{p},s)&=0~.
    \end{align} 
\end{subequations}
We note the useful identities
\begin{subequations}
\begin{align}
    m\xi_+-\mu\xi_-&=+M\xi_+~,\\
    m\xi_--\mu\xi_+&=-M\xi_-~,\\
    \xi_+^2-\xi_-^2&=1~,\\
    (\xi_+^2+\xi_-^2)M&=m~,\\
    2\xi_+\xi_-M&=\mu~.
\end{align}
\end{subequations}

The above expressions for the four-spinors in the Dirac basis can be obtained from the chiral basis, in terms of the two-component spinors, by means of the usual basis transformation
\begin{subequations}
\begin{align}
    u(\mathbf{p},s)=\frac{1}{\sqrt{2}}\begin{pmatrix} \mathbb{I}_2 & \mathbb{I}_2 \\ -\mathbb{I}_2 & \mathbb{I}_2\end{pmatrix}\begin{pmatrix} x(\mathbf{p},s) \\ \tilde{y}^{\dag}(\mathbf{p},s)\end{pmatrix}~,\\
    v(\mathbf{p},s)=\frac{1}{\sqrt{2}}\begin{pmatrix} \mathbb{I}_2 & \mathbb{I}_2 \\ -\mathbb{I}_2 & \mathbb{I}_2\end{pmatrix}\begin{pmatrix} y(\mathbf{p},s) \\ \tilde{x}^{\dag}(\mathbf{p},s)\end{pmatrix}~,
\end{align}
\end{subequations}
after making use of the identities (see Ref.~\cite{Dreiner:2008tw})
\begin{subequations}
\label{eq:sqrt_sig_p_identities}
\begin{align}
    \sqrt{\sigma\cdot p}=\frac{1}{\sqrt{2}}\left[(E+M)^{1/2}\mathbb{I}_2-(E+M)^{-1/2}\bm{\sigma}\cdot\mathbf{p}\right]~,\\
    \sqrt{\bar{\sigma}\cdot p}=\frac{1}{\sqrt{2}}\left[(E+M)^{1/2}\mathbb{I}_2+(E+M)^{-1/2}\bm{\sigma}\cdot\mathbf{p}\right]~,
\end{align}
\end{subequations}
and the property
\begin{equation}
    \bm{\sigma}\cdot \mathbf{p}u_s(\mathbf{s})=s|\mathbf{p}|u_s(\mathbf{s})~.
\end{equation}

Given the explicit forms of the four spinors, we can confirm that
\begin{subequations}
\label{eq:uvorthog}
 \begin{gather}
     u^{\dag}(\mathbf{p},s)\gamma^0u(\mathbf{p},s')=-v^{\dag}(\mathbf{p},s)\gamma^0v(\mathbf{p},s')=2M \, \delta_{ss'}~,\\
     u^{\dag}(\mathbf{p},s)\gamma^0v(\mathbf{p},s')=v^{\dag}(\mathbf{p},s)\gamma^0u(\mathbf{p},s')=0~,
 \end{gather}
\end{subequations}
and similarly for the tilded spinors, where we have used
\begin{equation}
u^{\dag}_s(\mathbf{s})u_{s'}(\mathbf{s})=\delta_{ss'}~.
\end{equation}
In addition, we have
\begin{subequations}
\begin{align}
    \sum_su(\mathbf{p},s)\tilde{u}^{\dag}(\mathbf{p},s)\, \gamma^0=\slashed{p}+m-\mu\gamma^5~,\\
    \sum_sv(\mathbf{p},s)\tilde{v}^{\dag}(\mathbf{p},s) \, \gamma^0=\slashed{p}-m+\mu\gamma^5~.
\end{align}
\end{subequations}

The canonical-conjugate $c$-number fields are $\psi$ and $\tilde{\psi}^{\dag}$, and we therefore introduce the quantum fields
\begin{subequations}
 \begin{align}
     \hat{\psi}(x)=\sum_{s=\pm}\int_{\mathbf{p}}\frac{1}{\sqrt{2E_{\mathbf{p}}}}\left[\hat{b}_{\mathbf{p},s}(0)u(\mathbf{p},s)e^{-ip\cdot x}+\hat{d}^{\dag}_{\mathbf{p},s}(0)v(\mathbf{p},s)e^{ip\cdot x}\right]~,\\
     \check{\psi}(x)=\sum_{s=\pm}\int_{\mathbf{p}}\frac{1}{\sqrt{2E_{\mathbf{p}}}}\left[\hat{b}_{\mathbf{p},s}(0)\tilde{u}(\mathbf{p},s)e^{-ip\cdot x}+\hat{d}^{\dag}_{\mathbf{p},s}(0)\tilde{v}(\mathbf{p},s)e^{ip\cdot x}\right]~,
 \end{align}
\end{subequations}
along with their Hermitian conjugates. Their evolution is governed by the Lagrangian [cf.~Eq.~\eqref{eq:DiracLag}]
\begin{equation}
    \label{eq:DiracLaghat}
    \hat{\mathcal{L}}
    =\check{\psi}^{\dag}\gamma^0i\slashed{\partial}\hat{\psi}-m\check{\psi}^{\dag}\gamma^0\hat{\psi}-\mu\check{\psi}^{\dag}\gamma^0\gamma^5\hat{\psi}
\end{equation}
and its Hermitian conjugate, which again yield mutually consistent Euler-Lagrange equations without further prescription.

The canonical equal-time anticommutation relations are
\begin{equation}
    \left\{\hat{\psi}(t,\mathbf{x}),\check{\psi}^{\dag}(t,\mathbf{y})\right\}=\left\{\check{\psi}(t,\mathbf{x}),\hat{\psi}^{\dag}(t,\mathbf{y})\right\}=\delta^3(\mathbf{x}-\mathbf{y})~,
\end{equation}
and we remark that the relevant propagators are those involving $\hat{\psi}$ and $\check{\psi}^{\dag}$. For example, the Feynman propagator is
\begin{equation}
    iS_{\rm F}(x,y)\equiv \braket{0|\mathrm{T}\left[\hat{\psi}(x)\check{\psi}^{\dag}(y)\gamma^0\right]|0}=i\int\frac{{\rm d}^4p}{(2\pi)^4}\,e^{-ip\cdot(x-y)}\frac{\slashed{p}+m-\mu\gamma^5}{p^2-M^2+i0^+}~,
\end{equation}
wherein $\mathrm{T}$ indicates time ordering.


\section{Matrix model}
\label{sec:matrix}

As we did for the bosonic theory in Ref.~\cite{AEMB}, we can construct a convenient matrix model that captures the salient non-Hermitian features of this fermionic theory. The matrix model of interest has Hamiltonian
\begin{equation}
\label{fermionmatrix}
    H=\begin{pmatrix} 0 & m+\mu \\ m-\mu & 0\end{pmatrix}~,
\end{equation}
reflecting the structure
\begin{equation}
m\gamma^0+\mu\gamma^0\gamma^5
\end{equation}
of the mass term of the field theory in the chiral basis. The eigenvectors of the Hamiltonian~\eqref{fermionmatrix} are 
\begin{equation}
    \mathbf{e}_{\pm}=N\begin{pmatrix} \pm\sqrt{1+\xi} \\ \sqrt{1-\xi}\end{pmatrix}~,
\end{equation}
where
\begin{equation}
\xi\equiv \frac{\mu}{m}    
\end{equation}
is the non-Hermitian parameter, and we take the normalization factor $N$ to be
\begin{equation}
    N=\frac{1}{\sqrt{2}\sqrt[4]{1-\xi^2}}~.
\end{equation}
As in the scalar case, the eigenvectors $\mathbf{e}_{\pm}$ are not orthogonal with respect to the Hermitian inner product:
\begin{equation}
    \mathbf{e}_{\pm}^*\cdot \mathbf{e}_{\mp}=-2N^2 \xi~.
\end{equation}

The parity matrix of this model (reflecting the parity transformation of the field theory) is
\begin{equation}
    P=\begin{pmatrix} 0 & 1 \\ 1 & 0\end{pmatrix}~,
\end{equation}
and we find that the eigenvectors are orthogonal with respect to the $\mathcal{PT}$ inner product:
\begin{equation}
    \mathbf{e}_{\pm}^{\mathcal{PT}}\cdot \mathbf{e}_{\mp}\equiv \mathbf{e}_{\pm}^{*}P\mathbf{e}_{\mp}=0~.
\end{equation}
However, one of the eigenvectors has a positive norm and the other a negative norm with respect to the $\mathcal{PT}$ inner product:
\begin{equation}
    \mathbf{e}_{\pm}^{\mathcal{PT}}\cdot \mathbf{e}_{\pm}=\pm 1~,
\end{equation}
as is expected for a non-Hermitian theory.

The Hamiltonian \eqref{fermionmatrix} is diagonalized by the similarity transformation
\begin{equation}
    H_{\rm diag}=S HS^{-1}~,
\end{equation}
where
\begin{equation}
\label{fermionR}
    S=N_S\begin{pmatrix} \sqrt{1-\xi} & \sqrt{1+\xi} \\ -\sqrt{1-\xi} & \sqrt{1+\xi}\end{pmatrix}~.
\end{equation}
The normalization factor $N_S$ is fixed below,
giving the Hermitian Hamiltonian
\begin{equation}
    H_{\rm diag}=m\sqrt{1-\xi^2}\begin{pmatrix} 1 & 0 \\ 0 &-1\end{pmatrix}~.
\end{equation}
However, since we have in mind a transformation to a Hermitian theory of a single Dirac fermion with squared mass $M^2=m^2-\mu^2$, we actually need to rotate this transformation through $\pi/4$, so that the non-zero entries lie in the elements associated with the operators $\bar{\psi}_L\psi_R$ and $\bar{\psi}_R\psi_L$, where $\psi_L$ and $\psi_R$ are the left- and right-chiral components of the Dirac field. We therefore define
\begin{equation}
    R=\frac{1}{\sqrt{2}}\begin{pmatrix} 1 & -1 \\ 1 & 1 \end{pmatrix} S~,
\end{equation}
giving the similarity transformation
\begin{equation}
    \label{eq:DiracHtoh}
    h=RHR^{-1}=\begin{pmatrix} 0 & m\sqrt{1-\xi^2} \\ m\sqrt{1-\xi^2} & 0\end{pmatrix}~.
\end{equation}
Choosing the normalization $N_S$ in Eq.~\eqref{fermionR} such that
\begin{equation}
    PR^{-1}P=R~,
\end{equation}
we obtain
\begin{equation}
    \label{eq:Rdef}
    R=\begin{pmatrix}\sqrt[4]{\frac{1-\xi}{1+\xi}} & 0 \\ 0 & \sqrt[4]{\frac{1+\xi}{1-\xi}} \end{pmatrix}~.
\end{equation}

The matrix $R$, along with the parity matrix $P$, can be used to construct an additional matrix $C'$, 
given by
\begin{equation}
    \label{eq:Cprimematrixmodel}
    C'=RPR^{-1}=\frac{1}{\sqrt{1-\xi^2}}\begin{pmatrix} 0 & 1-\xi \\ 1+\xi & 0 \end{pmatrix}~,
\end{equation}
which plays a key role in defining the positive-definite norm for the eigenstates of the non-Hermitian matrix~\eqref{fermionmatrix}.
We use it to construct the $\mathcal{C}'\mathcal{PT}$ inner product,  with respect to which the eigenvectors are orthonormal:
\begin{equation}
    \mathbf{e}_{\pm}^{\mathcal{C}'\mathcal{PT}}\cdot \mathbf{e}_{\pm}\equiv \mathbf{e}_{\pm}^{*}C'P\mathbf{e}_{\pm}=1~,\qquad 
    \mathbf{e}_{\pm}^{\mathcal{C}'\mathcal{PT}}\cdot \mathbf{e}_{\mp}=0~.
\end{equation}
The $C'$ matrix can also be used to define an additional matrix $Q$, which plays a role in the similarity transformation that maps this non-Hermitian theory to an equivalent Hermitian one (see, e.g., Refs.~\cite{Mostafazadeh:2001jk, BJR}). Specifically, we can write
\begin{equation}
    \label{eq:Dirac_sim}
    e^{-Q}=C'P=RPR^{-1}P=R^2=\frac{1}{\sqrt{1-\xi^2}}\begin{pmatrix} 1-\xi & 0 \\ 0 & 1+\xi\end{pmatrix}~,
\end{equation}
leading to
\begin{equation}
    \label{eq:DiracQmatrix}
    Q=\ln R^{-2}={\rm arctanh}\left(\xi\right)\begin{pmatrix}1 & 0 \\ 0 & -1 \end{pmatrix}~.
\end{equation}
It is interesting to consider the similarities between the form of this transformation and that of the scalar non-Hermitian field theory considered in Ref.~\cite{AEMB}.


\section{Similarity transformation}
\label{sec:similarity}

We saw in Section~\ref{sec:matrix} that the non-Hermitian matrix can be diagonalized via a similarity transformation. We emphasise that this transformation is not unitary.  The diagonalized matrix is Hermitian, and it is known that non-Hermitian theories can, in their regimes of unbroken antilinear symmetry, also be mapped to equivalent Hermitian theories via similarity transformations. In this Section, due to the fact that the non-Hermiticity of our prototype model resides only in the $c$-number spinor structure, we show that this transformation is simply a field redefinition. Nevertheless, we are able to construct an operator-valued expression for the similarity transform, which we relate to the discrete $\mathcal{C'}$ transformation in Section~\ref{sec:Fock}.

By inspection of Eq.~\eqref{eq:twospinors}, we see that by redefining
\begin{subequations}
\begin{align}
x_{\alpha}&\equiv\sqrt[4]{\frac{m+\mu}{m-\mu}}X_{\alpha}~,\qquad
&y_{\alpha}&\equiv\sqrt[4]{\frac{m+\mu}{m-\mu}}Y_{\alpha}~,\\
\tilde{x}^{\dag\dot{\alpha}}&\equiv\sqrt[4]{\frac{m-\mu}{m+\mu}}X^{\dag\dot{\alpha}}~,\qquad
&\tilde{y}^{\dag\dot{\alpha}}&\equiv\sqrt[4]{\frac{m-\mu}{m+\mu}}Y^{\dag\dot{\alpha}}~,\\
\hat{\chi}_{i,\alpha}&\equiv\sqrt[4]{\frac{m+\mu}{m-\mu}}\hat{\lambda}_{i,\alpha}~,\qquad
&\check{\chi}^{\dag\dot{\alpha}}_{i}&\equiv\sqrt[4]{\frac{m-\mu}{m+\mu}}\hat{\lambda}^{\dag\dot{\alpha}}_{i}~,
\end{align}
\end{subequations}
we arrive immediately at the Hermitian theory with Lagrangian
\begin{equation}
\label{eq:LDiracHermitian}
\hat{\mathcal{L}}'=i\hat{\lambda}^{\dag}_{i,\dot{\alpha}}\bar{\sigma}^{\nu\dot{\alpha}\beta}\partial_{\nu}\hat{\lambda}_{i,\beta}-m\sqrt{1-\xi^2}\hat{\lambda}_1^{\alpha}\hat{\lambda}_{2,\alpha}-m\sqrt{1-\xi^2}\hat{\lambda}_{2,\dot{\alpha}}^{\dag}\hat{\lambda}_1^{\dag\dot{\alpha}}~,
\end{equation}
where $\xi=\mu/m$ and
\begin{subequations}
\begin{align}
\hat{\lambda}_{1,\alpha}(x)=\sum_{s=\pm}\int_{\mathbf{p}}\frac{1}{\sqrt{2E_{\mathbf{p}}}}\left[\hat{d}_{\mathbf{p},s}(0)X_{\alpha}(\mathbf{p},s)e^{-ip\cdot x}+\hat{b}^{\dag}_{\mathbf{p},s}(0)Y_{\alpha}(\mathbf{p},s)e^{ip\cdot x}\right]~,\\
\hat{\lambda}_{2,\alpha}(x)=\sum_{s=\pm}\int_{\mathbf{p}}\frac{1}{\sqrt{2E_{\mathbf{p}}}}\left[\hat{b}_{\mathbf{p},s}(0)X_{\alpha}(\mathbf{p},s)e^{-ip\cdot x}+\hat{d}^{\dag}_{\mathbf{p},s}(0)Y_{\alpha}(\mathbf{p},s)e^{ip\cdot x}\right]~.
\end{align}
\end{subequations}
The triviality of the similarity transformation in the Weyl basis, when expressed in terms of action on the creation and annihilation operators, is a consequence of the fact that the similarity transformation does not mix the two Weyl spinors, resulting only in a straightforward rescaling.

The operator implementation of the transformation was given in Ref.~\cite{AEM4}, and takes the form
\begin{equation}
    \hat{\mathcal{L}}\to \hat{\mathcal{L}}'=\hat{\mathcal{S}}\hat{\mathcal{L}}\hat{\mathcal{S}}^{-1}~,
\end{equation}
with
\begin{equation}
    \hat{\mathcal{S}}=\exp\left[-\frac{1}{2}{\rm arctanh}\,\xi \int_{\mathbf{x}}\left(\check{\chi}_1^{\dag}(t,\mathbf{x})\hat{\chi}_1(t,\mathbf{x})+\check{\chi}_2^{\dag}(t,\mathbf{x})\hat{\chi}_2(t,\mathbf{x})\right)\right]~.
\end{equation}
Making use of
\begin{subequations}
\begin{align}
\int_{\mathbf{y}}\left[\check{\chi}^{\dag}_i(t,\mathbf{y})\hat{\chi}_i(t,\mathbf{y}),\hat{\chi}_i(t,\mathbf{x})\hat{\chi}_j(t,\mathbf{x})\right]&=-\left(1+\delta_{ij}\right)\hat{\chi}_i(t,\mathbf{x})\hat{\chi}_j(t,\mathbf{x})~,\\
\int_{\mathbf{y}}\left[\check{\chi}^{\dag}_i(t,\mathbf{y})\hat{\chi}_i(t,\mathbf{y}),\check{\chi}^{\dag}_j(t,\mathbf{x})\check{\chi}^{\dag}_i(t,\mathbf{x})\right]&=+\left(1+\delta_{ij}\right)\check{\chi}_j^{\dag}(t,\mathbf{x})\check{\chi}_i^{\dag}(t,\mathbf{x})~,\\
\int_{\mathbf{y}}\left[\check{\chi}_i^{\dag}(t,\mathbf{y})\hat{\chi}_i(t,\mathbf{y}),\check{\chi}_j^{\dag}(t,\mathbf{x})\bar{\sigma}\cdot \partial \hat{\chi}_j(t,\mathbf{x})\right]&=0~,
\end{align}
\end{subequations}
(see Ref.~\cite{AEM4}) we recover Eq.~\eqref{eq:LDiracHermitian}, but with $\lambda_i\to\chi_i$. If we instead try to write the transformation such that we recover precisely Eq.~\eqref{eq:LDiracHermitian}, as written in terms of $\lambda_i$, the transformation becomes trivial.

In the four-component basis, the rescaling takes the form
\begin{equation}
    \label{eq:Dirac_field_rescaling}
    \hat{\psi}\equiv\begin{pmatrix} \sqrt[4]{\frac{1+\xi}{1-\xi}} & 0 \\ 0 & \sqrt[4]{\frac{1-\xi}{1+\xi}} \end{pmatrix}\hat{\Psi}~,\qquad
    \check{\psi}^{\dag}\equiv \hat{\Psi}^{\dag}\begin{pmatrix} \sqrt[4]{\frac{1-\xi}{1+\xi}} & 0 \\ 0 & \sqrt[4]{\frac{1+\xi}{1-\xi}} \end{pmatrix}~,
\end{equation}
where we have suppressed two-dimensional unit matrices in the block form and
\begin{equation}
    \Psi=\begin{pmatrix} \hat{\lambda}_2 \\ \hat{\lambda}_1^{\dag} \end{pmatrix}~.
\end{equation}
We notice that the $4\times 4$ matrices involved in this rescaling are nothing but $R^{-1}\otimes \mathbb{I}_2$ and $R\otimes \mathbb{I}_2$, where $R$ was defined for the matrix model in Eq.~\eqref{eq:Rdef}. We see, by virtue of Eq.~\eqref{eq:Dirac_field_rescaling} and the non-unitary nature of the similarity transformation, the necessity to introduce the two types of field operators, viz.~the hatted and checked field operators.

Using the explicit form of the $\gamma$ matrices in the Weyl basis, i.e.,
\begin{equation}
    \gamma^0=\begin{pmatrix} 0 & \mathbb{I}_2 \\ \mathbb{I}_2 & 0 \end{pmatrix}, \qquad \gamma^5=\begin{pmatrix} -\mathbb{I}_2 & 0 \\ 0 & \mathbb{I}_2\end{pmatrix}~,
\end{equation}
we can show that
\begin{equation}
    \label{eq:matrixsimilarityDirac}
    \begin{pmatrix} \sqrt[4]{\frac{1-\xi}{1+\xi}} & 0 \\ 0 & \sqrt[4]{\frac{1+\xi}{1-\xi}} \end{pmatrix}\left(m\gamma^0+\mu\gamma^0\gamma^5\right)\begin{pmatrix} \sqrt[4]{\frac{1+\xi}{1-\xi}} & 0 \\ 0 & \sqrt[4]{\frac{1-\xi}{1+\xi}} \end{pmatrix}=m\sqrt{1-\xi^2}\gamma^0~.
\end{equation}
We then obtain the Hermitian Lagrangian directly:
\begin{equation}
    \hat{\mathcal{L}}'=\hat{\Psi}^{\dag}i\gamma^0\slashed{\partial}\hat{\Psi}-M\hat{\Psi}^{\dag}\gamma^0\hat{\Psi}~,
\end{equation}
where $M=\sqrt{m^2-\mu^2}$. The four-component field operators $\hat{\Psi}$ and $\hat{\Psi}^{\dag}$ are now built out of the usual four-spinors
\begin{subequations}
\label{eq:Hermitian_fourspinors}
\begin{align}
U(\mathbf{p},s)&=\begin{pmatrix} \sqrt{E_{\mathbf{p}}+M} \, u_s(\mathbf{s})\\ s\sqrt{E_{\mathbf{p}}-M} \, u_s(\mathbf{s})\end{pmatrix}~,\\
V(\mathbf{p},s)&=\begin{pmatrix} \sqrt{E_{\mathbf{p}}-M} \, u_{-s}(\mathbf{s})\\ -s\sqrt{E_{\mathbf{p}}+M} \, u_{-s}(\mathbf{s})\end{pmatrix}~.
\end{align}
\end{subequations}

If one were to insist on working with an operator transformation, things would become significantly more complicated. The Lagrangian \eqref{eq:DiracLaghat} can be mapped to a Hermitian one by the following similarity transformation, as first described in Ref.~\cite{BJR}:
\begin{equation}
    \hat{\mathcal{L}}\to \hat{\mathcal{L}}'=\hat{\mathcal{S}}\hat{\mathcal{L}}\hat{\mathcal{S}}^{-1}~,
\end{equation}
where
\begin{equation}
    \hat{\mathcal{S}}=e^{-\hat{\mathcal{Q}}/2}~,
\end{equation}
with
\begin{equation}
    \label{eq:QDirac}
    \hat{\mathcal{Q}}=-{\rm arctanh}\,\xi\int_{\mathbf{x}}\check{\psi}^{\dag}(t,\mathbf{x})\gamma^5\hat{\psi}(t,\mathbf{x})~,
\end{equation}
wherein we see the close analogy to the transformation of the matrix model in Eqs.~\eqref{eq:Dirac_sim} and \eqref{eq:DiracQmatrix}. The fields transform under this transformation as
\begin{subequations}
\begin{align}
    \hat{\psi}&\to\left(\cosh\frac{{\rm arctanh}\,\xi}{2}-\gamma^5\sinh\frac{{\rm arctanh}\,\xi}{2}\right)\hat{\psi}~,\\
    \check{\psi}^{\dag}&\to\check{\psi}^{\dag}\left(\cosh\frac{{\rm arctanh}\,\xi}{2}+\gamma^5\sinh\frac{{\rm arctanh}\,\xi}{2}\right)~.
\end{align}
\end{subequations}
Using the identities
\begin{subequations}
\begin{align}
    \cosh\frac{{\rm arctanh}\,\xi}{2}&=\frac{1}{\sqrt{2}}\sqrt{1+\frac{1}{\sqrt{1-\xi^2}}}~,\\
    \sinh\frac{{\rm arctanh}\,\xi}{2}&=\frac{1}{\sqrt{2}}\frac{\xi}{\sqrt{1-\xi^2}}\frac{1}{\sqrt{1+\frac{1}{\sqrt{1-\xi^2}}}}~,
\end{align}
\end{subequations}
we obtain after some algebra
\begin{equation}
    \hat{\mathcal{L}}'=\check{\psi}^{\dag}\gamma^0i\slashed{\partial}\hat{\psi}-M\check{\psi}^{\dag}\gamma^0\hat{\psi}~.
\end{equation}
This follows immediately from Eq.~\eqref{eq:matrixsimilarityDirac}, upon showing that
\begin{equation}
    \cosh\frac{{\rm arctanh}\,\xi}{2}\,\mathbb{I}_4-\sinh\frac{{\rm arctanh}\,\xi}{2}\,\gamma^5=\xi_+\mathbb{I}_4-\xi_-\gamma^5= \begin{pmatrix} \sqrt[4]{\frac{1+\xi}{1-\xi}} & 0 \\ 0 & \sqrt[4]{\frac{1-\xi}{1+\xi}} \end{pmatrix}~.
\end{equation}
For $0 \leq \xi < 1$, the operator transformation above can be written in terms of $\xi$, $\xi_+$ or $\xi_-$ by making use of the identities
\begin{equation}
    {\rm arctanh}\,\xi= 2\,{\rm arccosh}\,\xi_+=2\,{\rm arcsinh}\,\xi_-~.
\end{equation}
Notice, however, that the transformation maps the form of the Lagrangian, but not the field operators themselves (cf.~Ref.~\cite{AEMB}). Were we to try to construct a similarity transformation that maps both the Lagrangian and the fields, we would find trivial results, since the model is again a field rescaling away from Hermitian [see Eq.~\eqref{eq:Dirac_field_rescaling}].


\section{Discrete transformations in Fock space}
\label{sec:Fock}

In this Section, we discuss the discrete symmetry transformations in Fock space, namely the spacetime symmetry transformations of parity and time-reversal, and the $\mathcal{C}'$ transformation that arises in $\mathcal{PT}$-symmetric non-Hermitian theories.

\subsubsection{Parity}
\label{sec:parity}

Under a parity transformation, the spatial coordinates $\mathbf{x}$ change sign, i.e., $\mathbf{x}\to \mathbf{x}'=-\mathbf{x}$, but the time coordinate $t$ is unaffected, i.e.,
\begin{equation}
x^{\mu} \equiv (t,\mathbf{x})\to \mathcal{P}x^{\mu}=x^{\prime\mu}= (t^{\prime},\mathbf{x}^{\prime})=(t,-\mathbf{x})~.
\end{equation}
As a result, the three-momentum changes sign under parity, as does the helicity. For our non-Hermitian Dirac model, whose Lagrangian is \emph{not} invariant under parity, the Dirac fermion field operator transforms as
\begin{equation}
    \hat{\mathcal{P}}\hat{\psi}(t,\mathbf{x})\hat{\mathcal{P}}^{-1}=\gamma^0\check{\psi}(t,-\mathbf{x})~,
\end{equation}
where we emphasize that parity relates the hatted and checked operators. This follows from the transformation properties of the creation and annihilation operators
\begin{subequations}
\begin{align}
    \hat{\mathcal{P}}\hat{b}_{\mathbf{p},s}(0)\hat{\mathcal{P}}^{-1}=-s\hat{b}_{-\mathbf{p},-s}(0)~,\\
    \hat{\mathcal{P}}\hat{d}^{\dag}_{\mathbf{p},s}(0)\hat{\mathcal{P}}^{-1}=+s\hat{d}^{\dag}_{-\mathbf{p},-s}(0)~,
\end{align}
\end{subequations}
and the identities
\begin{subequations}
\begin{align}
    u(-\mathbf{p},-s)=-s\gamma^0\tilde{u}(\mathbf{p},s)~,\\
    v(-\mathbf{p},-s)=+s\gamma^0\tilde{v}(\mathbf{p},s)~,
\end{align}
\end{subequations}
where we have used the fact that
\begin{equation}
    u_s(-\mathbf{s})=su_{-s}(\mathbf{s}).
\end{equation}
As described for the bosonic case in Ref.~\cite{AEMB}, the definition of the parity transformation does not depend on the inner product used to define matrix elements, viz.~the Hermitian, $\mathcal{PT}$ or $\mathcal{C}'\mathcal{PT}$ inner product.

\subsubsection{Time-reversal}
\label{sec:treversal}

Under time reversal, the time coordinate $t$ changes sign, i.e., $t\to t'=-t$, but the spatial coordinates $\mathbf{x}$ are unaffected, i.e.,
\begin{equation}
x^{\mu} \equiv (t,\mathbf{x})\to \mathcal{T}x^{\mu}=x^{\prime\mu}= (t^{\prime},\mathbf{x}^{\prime})=(-t,\mathbf{x})~.
\end{equation}
The three-momentum also changes sign, but the helicity does not.

As identified in Ref.~\cite{AEMB}, in spite of the fact that the definition of the time-reversal operator in Fock space depends explicitly on the inner product used to define the matrix elements of the theory, its definition does not depend on whether we used the Hermitian, $\mathcal{PT}$ or $\mathcal{C}'\mathcal{PT}$ inner product, so the usual definitions of the time-reversal operator hold. Specifically, the Dirac fermion transforms as
\begin{subequations}
\begin{align}
    \hat{\mathcal{T}}\hat{\psi}(t,\mathbf{x})\hat{\mathcal{T}}^{-1}&=i\gamma^1\gamma^3\hat{\psi}(-t,\mathbf{x})~,\\
    \hat{\mathcal{T}}\check{\psi}^{\dag}(t,\mathbf{x})\hat{\mathcal{T}}^{-1}&=\check{\psi}^{\dag}(-t,\mathbf{x})i\gamma^1\gamma^3~.
\end{align}
\end{subequations}
As for the case of parity, this follows from the transformations of the creation and annihilation operators
\begin{subequations}
\begin{align}
    \hat{\mathcal{T}}\hat{b}_{\mathbf{p},s}(0)\hat{\mathcal{T}}^{-1}=\hat{b}_{-\mathbf{p},s}(0)~,\\
    \hat{\mathcal{T}}\hat{d}^{\dag}_{\mathbf{p},s}(0)\hat{\mathcal{T}}^{-1}=\hat{d}^{\dag}_{-\mathbf{p},s}(0)~,
\end{align}
\end{subequations}
and the identities
\begin{subequations}
\begin{align}
    u^*(-\mathbf{p},s)=i\gamma^1\gamma^3u(\mathbf{p},s)~,\\
    v^*(-\mathbf{p},s)=i\gamma^1\gamma^3v(\mathbf{p},s)~,
\end{align}
\end{subequations}
wherein we have made use of the additional relation
\begin{equation}
u_s^*(-\mathbf{s})=-i\sigma^1\sigma^3u_s(\mathbf{s})~.
\end{equation}

\subsubsection{$\hat{\mathcal{C}}'$ operator}
\label{sec:Cprime}

By introducing the tilded and untilded spinors, we have seen that this non-Hermitian Dirac model is a field redefinition away from being Hermitian. We have also seen that the non-Hermitian nature of the model resides only in the spinor structure, and the algebra of the creation and annihilation operators is just that of a Hermitian theory.

Consider now the operator
\begin{equation}
    \hat{\mathcal{O}}=e^{-\hat{\mathcal{Q}}}~,
\end{equation}
with $\hat{\mathcal{Q}}$ given by Eq.~\eqref{eq:QDirac}. This acts on the fields as
\begin{subequations}
\begin{align}
    \hat{\psi}&\to\left(\cosh\,{\rm arctanh}\,\xi-\gamma^5\sinh\,{\rm arctanh}\,\xi\right)\hat{\psi}~,\\
    \check{\psi}^{\dag}&\to\check{\psi}^{\dag}\left(\cosh\,{\rm arctanh}\,\xi+\gamma^5\sinh\,{\rm arctanh}\,\xi\right)~.
\end{align}
\end{subequations}
Using the identities
\begin{subequations}
\begin{align}
\cosh\,{\rm arctanh}\,\xi=\frac{1}{\sqrt{1-\xi^2}}~,\\
\sinh\,{\rm arctanh}\,\xi=\frac{\xi}{\sqrt{1-\xi^2}}~,
\end{align}
\end{subequations}
and the fact that
\begin{equation}
    \frac{1}{1-\xi^2}(\mathbb{I}_4+\xi\gamma^5)\gamma^0(m+\mu\gamma^5)(\mathbb{I}_4-\xi\gamma^5)=\gamma^0(m-\mu\gamma^5)~,
\end{equation}
we can convince ourselves that the action of $\hat{\mathcal{O}}$ does not leave the Hamiltonian invariant, as would be required for the $\hat{\mathcal{C}}'$ operator (see Ref.~\cite{Bender:2002vv}).

If we were to take inspiration from the quantum mechanics case (see Ref.~\cite{Bender:2002vv}), we would compose the operator $\hat{\mathcal{O}}$ with the parity operator $\hat{\mathcal{P}}$. This yields
\begin{subequations}
\begin{align}
    \hat{\mathcal{O}}\hat{\mathcal{P}}\hat{\psi}(t,\mathbf{x})\hat{\mathcal{P}}^{-1}\hat{\mathcal{O}}^{-1}&=\frac{1}{\sqrt{1-\xi^2}}\gamma^0(\mathbb{I}_4+\xi\gamma^5)\check{\psi}(t,-\mathbf{x})~,\\
    \hat{\mathcal{O}}\hat{\mathcal{P}}\check{\psi}^{\dag}(t,\mathbf{x})\hat{\mathcal{P}}^{-1}\hat{\mathcal{O}}^{-1}&=\frac{1}{\sqrt{1-\xi^2}}\hat{\psi}^{\dag}(t,-\mathbf{x})(\mathbb{I}_4-\xi\gamma^5)\gamma^0~.
\end{align}
\end{subequations}
While this leaves the factor $m+\mu\gamma^5$ invariant, i.e.,
\begin{equation}
    \frac{1}{1-\xi^2}(\mathbb{I}_4-\xi\gamma^5)(\gamma^0)^2(m+\mu\gamma^5)\gamma^0(\mathbb{I}_4+\xi\gamma^5)=\gamma^0(m+\mu\gamma^5)~,
\end{equation}
it simultaneously transforms $\hat{\psi}(t,\mathbf{x})\to\check{\psi}(t,-\mathbf{x})$,
and is therefore not a symmetry of the Hamiltonian.

In order to construct $\hat{\mathcal{C}}'$, we would therefore need to compose further with an operator $\hat{\mathcal{P}}_+$ that has the following action
on the fermion field:
\begin{subequations}
\begin{align}
    \hat{\mathcal{P}}_+\hat{\psi}(t,\mathbf{x})\hat{\mathcal{P}}^{-1}_+&=\check{\psi}(t,-\mathbf{x})~,\\
    \hat{\mathcal{P}}_+\check{\psi}^{\dag}(t,\mathbf{x})\hat{\mathcal{P}}^{-1}_+&=\hat{\psi}^{\dag}(t,-\mathbf{x})~,
\end{align}
\end{subequations}
without the appearance of the parity matrix $P=\gamma^0$. In this way, the operator $\hat{\mathcal{C}}'$ would take a form analogous to the scalar case reported in Ref.~\cite{AEMB}. Specifically (cf.~Ref.~\cite{BJR}),
\begin{equation}
    \hat{\mathcal{C}}'=e^{-\hat{\mathcal{Q}}}\hat{\mathcal{P}}\hat{\mathcal{P}}_+~.
\end{equation}
However, the existence of such a $\hat{\mathcal{P}}_+$ operator is an open question. Even so, its necessary appearance in the quantum field theory case, compared to the quantum mechanics case, can be understood as a consequence of the internal degrees of freedom of quantum field operators and the fact that single-particle Fock states are simultaneously eigenstates of the momentum and energy operators. On the other hand, for any quantum mechanical theory with a non-Hermitian potential, e.g., the theory with Hamiltonian $\hat{H}=\hat{p}^2+i\hat{x}^3$, eigenstates of the momentum operator $\hat{p}$ are not eigenstates of the Hamiltonian $\hat{H}$ given its dependence on the position operator $\hat{x}\equiv x$.


\section{Inner products in Fock space}
\label{sec:inner}

As noted previously, the non-Hermitian nature of this model resides only in the spinor structure. Single-particle momentum eigenstates $\ket{\mathbf{p},s}$ and $\ket{\mathbf{p}',s'}$, say, of momentum $\mathbf{p}$ and $\mathbf{p}'$, and helicities $s$ and $s'$, respectively, are therefore orthogonal with respect to the usual Hermitian inner product; namely,
\begin{equation}
    \left(\ket{\mathbf{p},s}\right)^{\dag}\ket{\mathbf{p}',s'}=(2\pi)^3\delta_{ss'}\delta^3(\mathbf{p}-\mathbf{p}')~.
\end{equation}
So long as we work with the true canonical conjugate field operators, $\hat{\psi}$ and $\check{\psi}^{\dag}$, we  will always obtain consistent combinations of tilded and untilded classical spinor factors. We therefore do not need to construct any additional inner products for this model.

It is interesting to note that the $\mathcal{PT}$ norm is not positive-definite for this model, as is expected for such non-Hermitian theories. The $\mathcal{PT}$ inner product of the single-particle states would give
\begin{equation}
        \left(\hat{\mathcal{P}}\hat{\mathcal{T}}\ket{\mathbf{p},s}\right)^{\mathsf{T}}\ket{\mathbf{p}',s'}=(2\pi)^3\delta_{(-s)s'}\delta^3(\mathbf{p}-\mathbf{p}')~.
\end{equation}
The change of sign on the helicity results from the fact that parity flips the helicity but time-reversal does not. We see then that the $\mathcal{PT}$ norm of the single-particle momentum states would be zero.


\section{Conclusions}
\label{sec:conx}

In this paper, we have extended our previous discussion of discrete spacetime symmetries
in bosonic quantum field theories~\cite{AEMB} to the case of a single Dirac fermion with four components, the minimum required to realize non-trivial $\mathcal{PT}$ symmetry.
We discussed its parity $\mathcal{P}$ and time-reversal $\mathcal{T}$ transformations at the classical level,
and the quantum version of the theory, revisiting its discrete symmetries including the
$\mathcal{C}'$ symmetry~\cite{Bender:2002vv}. We have shown that the Fock space of this model is that of a Hermitian theory, while the non-Hermiticity resides in the spinor structure. As a result, the inner product of single-particle momentum and helicity eigenstates is the usual Hermitian one. We have nevertheless emphasized that the $\mathcal{PT}$ inner product is still not positive-definite, as expected for a non-Hermitian $\mathcal{PT}$ symmetric theory.

We have established that this model is a field redefinition (in spinor space) away from being Hermitian and that this is most easily seen in the two-component basis.
Nevertheless, we have constructed the $\mathcal{C}'$ transformation under which the Hamiltonian is invariant and shown
how this is related to the similarity transformation that diagonalizes the corresponding Hamiltonian.

Our analysis carries forward the programme of establishing the consistency of $\mathcal{PT}$-symmetric
quantum field theories, which will entail many further steps. These include the extension to multiple flavours
and Majorana fermion models, and a more rigorous treatment of interactions
between fermions, scalars and gauge fields, which will require deeper understanding of the path integral
in such theories than is currently available and allow, e.g., to construct non-Hermitian extensions of the SM Higgs sector (see the differing approaches of Refs.~\cite{AEMS1, AEMS2, AEMS3} and Refs.~\cite{Mannheim, FT1, FT2, Fring:2020bvr}). Our motivation in pursuing this programme is largely due to the
possibility that $\mathcal{PT}$-symmetric theories may offer generalizations of conventional quantum field
theories with interesting applications in fundamental physics, by offering a novel framework for new physics
beyond the Standard Model. We plan to return to these issues in future publications.


\section*{Acknowledgements}

PM would like to thank Maxim N.~Chernodub, Madeleine Dale, Robert Mason and Esra Sablevice for interesting discussions of related non-Hermitian theories. The work of JA and JE was supported by the United Kingdom Science and Technology Facilities Council (STFC) [Grant Nos.~ST/P000258/1 and ST/T000759/1] and Engineering and Physical Sciences Research Council (EPSRC) [Grant No.~EP/V002821/1], 
and that of JE also by the Estonian Research Council via a Mobilitas Pluss grant. The work of PM was supported by a Leverhulme Trust Research Leadership Award [Grant No.~RL-2016-028]; a Nottingham Research Fellowship from the University of Nottingham; and a United Kingdom Research and Innovation (UKRI) Future Leaders Fellowship [Grant No.~MR/V021974/1].


\end{document}